\documentclass[journal]{IEEEtranTCOM}
\normalsize

\usepackage[mathscr]{eucal}
\usepackage[cmex10]{amsmath}
\usepackage{epsfig,epsf,psfrag}
\usepackage{amssymb,amsmath,amsthm,amsfonts,latexsym}
\usepackage{amsmath,graphicx,bm,xcolor,url}
\usepackage[caption=false]{subfig} 
\usepackage{fixltx2e}
\usepackage{array}
\usepackage{verbatim}
\usepackage{bm}
\usepackage{algorithmic, cite}
\usepackage{algorithm}
\usepackage{verbatim}
\usepackage{textcomp}
\usepackage{mathrsfs}
\usepackage{multirow}
\usepackage{epstopdf}
\usepackage{multicol}
\usepackage{ifpdf}

\catcode`~=11 \def\UrlSpecials{\do\~{\kern -.15em\lower .7ex\hbox{~}\kern .04em}} \catcode`~=13 

\allowdisplaybreaks[1]

\newcommand{\calC}{\mathcal{C}}

\newcommand{\calN}{\mathcal{N}}

\newcommand{\calX}{\mathcal{X}}

\newcommand{\rmb}{\mathrm{b}}

\newcommand{\rme}{\mathrm{e}}

\newcommand{\rmk}{\mathrm{k}}

\newcommand{\rmx}{\mathrm{x}}

\newcommand{\bbE}{\mathbb{E}}

\newcommand{\bbP}{\mathbb{P}}

\newcommand{\bbR}{\mathbb{R}}

\newcommand{\bbZ}{\mathbb{Z}}

\DeclareMathAlphabet{\mathbsf}{OT1}{cmss}{bx}{n}
\DeclareMathAlphabet{\mathssf}{OT1}{cmss}{m}{sl}

\newcommand{\rvP}{\mathsf{P}}

\DeclareSymbolFont{bsfletters}{OT1}{cmss}{bx}{n}  
\DeclareSymbolFont{ssfletters}{OT1}{cmss}{m}{n}
\DeclareMathSymbol{\bsfGamma}{0}{bsfletters}{'000}
\DeclareMathSymbol{\ssfGamma}{0}{ssfletters}{'000}
\DeclareMathSymbol{\bsfDelta}{0}{bsfletters}{'001}
\DeclareMathSymbol{\ssfDelta}{0}{ssfletters}{'001}
\DeclareMathSymbol{\bsfTheta}{0}{bsfletters}{'002}
\DeclareMathSymbol{\ssfTheta}{0}{ssfletters}{'002}
\DeclareMathSymbol{\bsfLambda}{0}{bsfletters}{'003}
\DeclareMathSymbol{\ssfLambda}{0}{ssfletters}{'003}
\DeclareMathSymbol{\bsfXi}{0}{bsfletters}{'004}
\DeclareMathSymbol{\ssfXi}{0}{ssfletters}{'004}
\DeclareMathSymbol{\bsfPi}{0}{bsfletters}{'005}
\DeclareMathSymbol{\ssfPi}{0}{ssfletters}{'005}
\DeclareMathSymbol{\bsfSigma}{0}{bsfletters}{'006}
\DeclareMathSymbol{\ssfSigma}{0}{ssfletters}{'006}
\DeclareMathSymbol{\bsfUpsilon}{0}{bsfletters}{'007}
\DeclareMathSymbol{\ssfUpsilon}{0}{ssfletters}{'007}
\DeclareMathSymbol{\bsfPhi}{0}{bsfletters}{'010}
\DeclareMathSymbol{\ssfPhi}{0}{ssfletters}{'010}
\DeclareMathSymbol{\bsfPsi}{0}{bsfletters}{'011}
\DeclareMathSymbol{\ssfPsi}{0}{ssfletters}{'011}
\DeclareMathSymbol{\bsfOmega}{0}{bsfletters}{'012}
\DeclareMathSymbol{\ssfOmega}{0}{ssfletters}{'012}

\newcommand{\tilD}{\tilde{D}}

\newcommand{\tilf}{\tilde{f}}

\newcommand{\tilg}{\tilde{g}}

\newcommand{\tilh}{\tilde{h}}

\newcommand{\bone}{\mathbf{1}}

\newtheorem{theorem}{Theorem} 
\newtheorem{lemma}{Lemma}

\newtheorem{proposition}{Proposition}
\newtheorem{corollary}{Corollary}
\newtheorem{definition}{Definition}

\newtheorem{remark}{Remark}

\newcommand{\qednew}{\nobreak \ifvmode \relax \else
      \ifdim\lastskip<1.5em \hskip-\lastskip
      \hskip1.5em plus0em minus0.5em \fi \nobreak
      \vrule height0.75em width0.5em depth0.25em\fi}

\usepackage[T1]{fontenc}
\usepackage{flushend}
\usepackage{dsfont}
\usepackage{xspace}
\usepackage{graphicx}
\usepackage{pgfgantt}
\usepackage{float}
\usepackage{caption}
\usepackage[ colorlinks = true,
             linkcolor = blue,
             urlcolor  = blue,
             citecolor = red,
             anchorcolor = green,
]{hyperref}

\allowdisplaybreaks[2]
\begin{document}

\title{Performance of Viterbi Decoding with and without ARQ on Rician Fading Channels}
\author{Lan V.\ Truong, \emph{Member, IEEE}
\thanks{This paper was presented in part at 2018 IEEE International Symposium on Information Theory~\cite{Truong18e}.}
\thanks{The author is with the  Department of Computer Science, National University of Singapore (NUS). Email: \url{truongvl@comp.nus.edu.sg}.}} 
\markboth{IEEE Transactions on Communications}%
{Submitted paper}

\maketitle

\begin{abstract} In this paper, we investigate the performance of the Viterbi decoding algorithm with/without Automatic Repeat reQuest (ARQ) over a Rician flat fading channel with unlimited interleaving. We show that the decay rate of the average bit error probability with respect to the bit energy to noise ratio is at least equal to $d_f$ at high bit energy to noise ratio for both cases (with ARQ and without ARQ), where $d_f$ is the free distance of the convolutional code. The Yamamoto-Itoh flag helps to reduce the average bit error probability by a factor of $4^{d_f}$ with a negligible retransmission rate. We also prove an interesting result that the average bit error probability decays exponentially fast with respect to the Rician factor for any fixed bit energy per noise ratio. In addition, the average bit error exponent with respect to the Rician factor is shown to be $d_f$. 
\end{abstract}   
\begin{IEEEkeywords}
Coded ARQ, Hybrid ARQ, Feedback, Rician Fading, Error Exponents, Block Fading Channels.
\end{IEEEkeywords}
\section{Introduction}
Viterbi,\cite{Viterbi1967}, proposed a non-sequential decoding algorithm and derived an upper bound on error probability using random coding arguments for Discrete Memoryless (DM) and Additive White Gaussian Noise (AWGN) channels. The algorithm was thereafter shown to yield maximum likelihood decisions by Omura,\cite{Omura1969}, and Forney~\cite{Forney1974}. In 1971, Viterbi proposed a method to evaluate the error probabilities of convolutional codes by using their transfer functions~\cite{Viterbi1971}. The performance of the convolutional codes,\cite{Viterbi1967}, was later evaluated for time-varying channels. Using the same transfer function method, the error probability of the Viterbi decoding over the Rayleigh fading channel with unlimited interleaving was estimated by evaluating the exact pairwise error probabilities together with the union bound~\cite{Proakis}. Vucetic evaluated the performance of punctured convolutional codes with maximum-likelihood Viterbi algorithm to enable adaptive encoding and decoding without modifying the basic structure of the encoder and the decoder for Rician fading channels~\cite{Vucetic}. Malkamaki and Leib considered union upper bound techniques for convolutional codes with unlimited interleaving over block fading Rician channels~\cite{MalkamakiLeib}. Although their method provides useful numerical results, the performance is intractable (hard to analyze).

Automatic Repeat reQuest (ARQ),\cite{LinCostello}, is an error-control method for data transmission that uses acknowledgments to achieve reliable data transmission over an unreliable service. As the turn-around time of the communication link increases, however, retransmission becomes expensive and a more elaborate technique with reduced retransmission is required. Coded ARQ, which combines error-correcting coding and retransmission, is then an alternative to ARQ. Fang,\cite{Fang1971a}, and Yamamoto-Itoh,\cite{YamamotoItoh1980}, studied convolutionally coded ARQ schemes with Viterbi decoding,\cite{Viterbi1967}, and showed that a low error probability is attained by having a moderate increase in complexity for DM and AWGN channels. However, the frequency of retransmission in the Yamamoto-Itoh algorithm is much less than that of the Fang algorithm. In such schemes, the encoders with full knowledge of output sequences and receivers' decoding algorithms can estimate the receivers' decoded messages at each fixed interval and compare with the transmitted messages in order to decide whether to stop or to continue transmission after each fixed interval as in the coded-ARQ.

The Yamamoto-Itoh algorithm has been implemented in the Keystone Architecture Viterbi-Decoder Coprocessor (VCP2), Texas Instruments\cite{TexasIns}. A new decoding algorithm based on the modified Viterbi algorithm for repeat request systems was proposed in~\cite{Kudryashov1993}, which was numerically shown to achieve a better error exponent than Yamamoto-Itoh's result,\cite{YamamotoItoh1980}, for the binary symmetric channel with crossover probability $0.1$. The error exponent with respect to convolutional block length for different channel models was also considered in a number of papers,\cite{YamamotoItoh1980, Hashimoto1994,Yamamoto1995,Fujiwara1995,Raghavan1998,Hashimoto1999, Malkamaki199b}, by using Gallanger's error exponent~\cite{Gallager1965a}. Hashimoto,\cite{Hashimoto1994}, theoretically proved that the Yamamoto-Itoh algorithm attains a better error exponent than the one shown in\cite{YamamotoItoh1980} for DM channels. In~\cite{Raghavan1998}, the ROVA (Reliability Output Viterbi Algorithm) was proposed for hybrid-ARQ, and the performance was compared to the Yamamoto-Itoh algorithm. A combined scheme of the Yamamoto-Itoh algorithm and ROVA scheme was proposed in~\cite{Hashimoto1999}, and it was shown theoretically that the error exponent was improved compared to Yamamoto-Itoh algorithm for very noisy channels. Since no simulation results were shown, it is uncertain whether the combined scheme can actually attain better performance compared to the Yamamoto-Itoh algorithm for practical channels. There are some other related papers which proposed algorithms to improve performance of hybrid-ARQ~\cite{Chen2001,Stender2004,Pai2011}. For example, H. T. Pai et al. proposed an algorithm to reduce the retransmission probability when a transmission error appears in a large packet. Some papers consider performance of a combination of ARQ and other types of error control codes such as turbo codes~\cite{Makki2014} or LDPC codes~\cite{Soijanin2006}. 

In this paper, we investigate the performance of the original Viterbi decoder,\cite{Viterbi1967}, and the modified Viterbi decoding algorithm by Yamamoto-Itoh,\cite{YamamotoItoh1980}, over Rician fading channels with unlimited interleaving. The original Viterbi decoder,\cite{Viterbi1967}, can be considered as the Yamamoto-Itoh algorithm when setting the Yamamoto-Itoh flag equal to zero ($u=0$) \cite{YamamotoItoh1980}. We show that the decay rate of the average bit error probability of the original Viterbi decoding scheme,\cite{Viterbi1967}, is at least equal to $d_f$ at high bit energy to noise ratio $E_b/N_0$. In addition, there exists a Yamamoto-Itoh flag such that the bit error probability of the Yamamoto-Itoh algorithm is lowered by at least a factor of $4^{d_f}$. When the Rician factor becomes very large, the bit error probability is shown to decay exponentially fast with the convolutional code free-distance being its exponent. To the best of the author's knowledge, these results have not appeared in the literature before. All the existing results, thus far, were mainly evaluated for the original Viterbi decoding,\cite{Viterbi1967}, or the Yamamoto-Itoh algorithm,\cite{YamamotoItoh1980}, over DM or AWGN channels. For Rayleigh or Rician fading channels, the majority of works concentrate on providing numerical results for different channel situations. In addition, the important effects of Rician factor on the error exponent for a fixed bit energy to noise ratio have not been analytically expressed in closed forms.

The motivation for this work is two fold. Firstly, the Rician fading channel is a practical communication channel model in wireless communications when one of the paths, typically the line of sight signal, is much stronger than the others. Research literature has mainly focused on studying the effects of the signal to noise ratio on the bit error rate for communication over this channel, however the effects of the line of sight path (Rician factor) on performance are not fully understood. There has been no analytical result (in closed forms) which shows the effects of the Rician factor on the error performance when using some practical coding schemes for wireless communication channels. Secondly, there have been a recent research trend on reconsidering the effects of variable-lengh feedback codes on improving the error exponents and capacities for DM and AWGN channels~\cite{Burnashev1976, YamamotoItoh1979, Yury2011, TruongTan17BC,TruongTan18aMACVL}. However, how to design practical coding schemes to achieve optimal performances has been still unknown. The Yamamoto-Itoh coding algorithm,~\cite{YamamotoItoh1980}, is a simple practical variable-length coding scheme which uses only one bit feedback each round (ARQ) to (at least) double the reliability function for DM channels and also to improve the error performance on AWGN channels. However, the measured performance on in the Yamamoto-Itoh paper,~\cite{YamamotoItoh1980} is the reliability function, i.e., for sufficiently large code-length. An interesting open question is that whether the Yamamoto-Itoh coding scheme has some positive effects on the error performance at finite code-length or not, especially for some practical wireless communication channels?  We note that the performance of convolutional codes with modified Viterbi decoding algorithms at finite code block-length has been recently considered in~\cite{newMakki2014},~\cite{Williamson2014} using higher complexity decoding algorithms such as the ROVA and tail-biting ROVA. However, these works are mainly for DM or AWGN channels,\cite{Williamson2014}, or Rayleigh fading channels but from capacity perspective (outage capacity)~\cite{newMakki2014}. Some other interesting papers,~\cite{Makki2014e},~\cite{CaireARQ2001}, also consider affects of AQR on performance of wireless communication systems when combining with random coding schemes, i.e., from capacity viewpoints. The positive effects of ARQ on tradeoff between the error probability and retransmission probability for practical coding schemes for wireless communications have been still unknown.

The rest of this paper is organized as follows: The channel model is provided in Section~\ref{ch:model}. Some mathematical preliminaries are introduced in Section~\ref{math_pre}. Our main result is stated in Section~\ref{main}. The proof of this main result (Theorem~\ref{thm}) and numerical evaluations of tradeoffs between performance measures (bit error probability and retransmission probability) are provided in Section~\ref{sec:lower}. Proofs that are more technical are deferred to appendices. 
\section{Channel Model}\label{ch:model}
 \begin{figure}[H]
	\centering
		\includegraphics[width=0.5\textwidth]{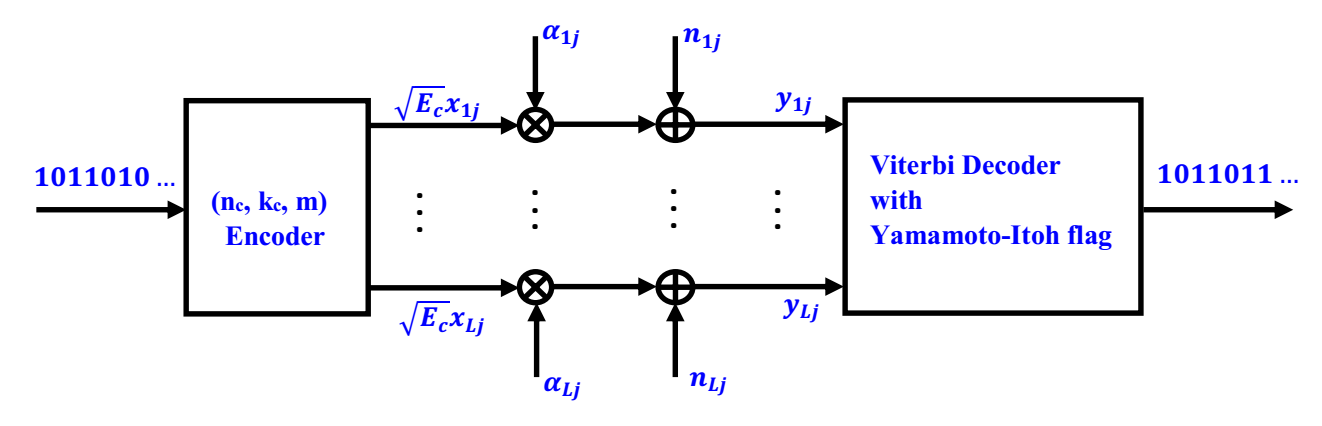}
	\caption{System Model}
	\label{fig:Channel}
\end{figure}
\subsection{Channel Models} We investigate a communication channel model in Fig.~\ref{fig:Channel} where the decoder uses the Yamamoto-Itoh algorithm for decoding the convolutional code~\cite{YamamotoItoh1980}. The Yamamoto-Itoh algorithm is a modification of the original Viterbi decoding,~\cite{Viterbi1967}, to make it work better for DM and AWGN channels with ARQ. A block of $H k_c$ bits from the data source for $H \in \bbZ^+$ are first encoded by a convolutional code of rate $R_c = k_c/n_c$ and with constraint length $k_c K$, where $K=m + 1$ and $m$ is the memory order of the code. Before encoding, $m k_c$ tail bits are added to each block of $H k_c$ bits to terminate the code trellis into a known state. The $n_c (H+m)$ encoder output bits are denoted by $x_{ij}$, where $i\in \{1,2,\ldots,n_c\}$ indicates the index of a column in the generator polynomial matrix associated with the convolutional code and $j\in \{1,2,\ldots,H+m\}$. In the analysis, we assume that antipodal modulation is performed, i.e., $x_{ij}=\pm 1$. The output bits $x_{ij}$ are interleaved over $L$ subchannels. To simplify the analysis, we assume that $L=n_c$.

The subchannels are assumed to be frequency nonselective fading Rician and independent of each other. In this model, the fading process is assumed to be constant over a block of $n_c$ channel symbols (coherence time). This assumption allows the receiver to be able to estimate channel state information, i.e., $\{\alpha_{ij}\}_{i=1,n_c}^{j=1,H+m}$, perfectly. Assuming coherent detection, the received signal samples can be written as
\begin{align}
\label{eqn:eqkey1}
y_{ij}=\sqrt{E_c}\alpha_{ij} x_{ij}+n_{ij},
\end{align}
where $i$ indicates the subchannel, $j$ is the sample within a subchannel, $E_c$ is the energy per transmitted code symbol, and $n_{ij}$'s are zero-mean white Gaussian noise samples with variance $N_0/2$. The average energy per transmitted bit can be easily shown to be equal to
\begin{align}
\label{eqeb}
E_b = \left(\frac{H+m}{H}\right)\left(\frac{1}{R_c}\right) E_c.
\end{align}
The fading envelopes $\alpha_{ij}$ of the $n_c$ subchannels involved in each decoding process are assumed to be independent of each other, identically distributed, and constant over a block of $n_c$ channel symbols. Here, $\alpha_{ij}$ are assumed to be Rician distributed with noncentrality parameter $s \geq 0$, scale parameter $\sigma > 0$, and the probability density function
\begin{align}
\label{eqkey2}
f(\alpha_{ij})=\frac{\alpha_{ij}}{\sigma^2} e^{-((\alpha_{ij}^2+s^2)/2\sigma^2)}I_0\left(\frac{\alpha_{ij} s}{\sigma^2}\right), \quad \alpha_{ij} \geq 0,
\end{align}
where $\bbE(\alpha_{ij}^2)=s^2+ 2\sigma^2$. The Rician factor $\gamma$ is defined as $\gamma=s^2/(2\sigma^2)$, and
\begin{align}
\label{definemod}
I_0(x)=\frac{1}{2\pi} \int_{-\pi}^{\pi} e^{x\cos \theta} d\theta.
\end{align}
\subsection{Decoding Algorithm}\label{decodalg}
The Viterbi decoder with repeated request proposed by Yamamoto-Itoh (Yamamoto-Itoh algorithm) in Fig.~\ref{fig:Channel} is used for the decoding of convolutional codes, which employs the samples $y_{ij}$ as well as the ideal channel state information (CSI), $\hat{\alpha}_i=\alpha_i$~\cite{YamamotoItoh1980}. The branch metrics are calculated as
\begin{align}
\lambda_j^{(r)}=\sum_{i=1}^{n_c} \alpha_{ij} x^{(r)}_{ij} y^{(r)}_{ij}, \quad j \in \{1,2,\ldots, H+m\},
\end{align} where $x_{ij}^{(r)} \in \{x_{ij}\}_{i=1,n_c}^{j=1,(H+m)}, y_{ij}^{(r)} \in \{y_{ij}\}_{i=1,n_c}^{j=1,(H+m)}$ are obtained by reading the elements of the coded symbol matrix $\{x_{ij}\}_{i=1,n_c}^{j=1,H+m}$ of size $n_c\times (H+m)$ column-by-column from top-to-down. In general $x_{ij}^{(r)}\neq x_{ij}$, however for the case $L=n_c$, it holds that $x_{ij}^{(r)}=x_{ij}$. Notice that on each branch in the trellis, the first bit comes from one subchannel, the second bit from another subchannel, etc. 

Note that the Viterbi Decoder with Yamamoto-Itoh flag is a modified version of the Yamamoto-Itoh Algorithm,~\cite{YamamotoItoh1980}, to deal with our fading channel model. This modified decoding algorithm is as follows. To begin with, at level $K-1$, put a label $\calC$ on all $2^{K-1}$ paths. At each node of level $t$ for $t\in \{K,K+1,K+2,\ldots,H+m\},$ the decoder estimates the sum of branch metrics $\sum_{j=1}^t\lambda_j^{(r)}$ by using dynamic programming as the original Viterbi decoding algorithm,~\cite{Viterbi1971}, and then selects two paths  $\{x_{ij}^{(r)}\}_{i=1,n_c}^{j=1,t}$ and $\{(x_{ij}^{(r)})'\}_{i=1,n_c}^{j=1,t}$ that have the largest sum of branch metrics $\sum_{j=1}^t\lambda_j^{(r)}$ and the second largest one, respectively. This means that after first round of choice, we have
\begin{align}
\label{cond1}
\sum_{j=1}^t \sum_{i=1}^{n_c} \alpha_{ij}x_{ij}^{(r)}y_{ij}  \geq \sum_{j=1}^t \sum_{i=1}^{n_c} \alpha_{ij} (x_{ij}^{(r)})'y_{ij}.
\end{align}
Now, if the path $\{x_{ij}^{(r)}\}_{i=1,n_c}^{j=1,t}$ has label $\calC$ at level $t-1$ and the following condition
\begin{align}
\label{eq7keykey}
&\sum_{j=1}^t \sum_{i=1}^{n_c} \alpha_{ij} \left(x_{ij}^{(r)}-(x_{ij}^{(r)})'\right) y_{ij} \nonumber\\
&\qquad \geq \frac{u}{d_f}  \sqrt{N_0/2}\sum_{j=1}^t \sum_{i=1}^{n_c}\left|(x_{ij}^{(r)})'-x_{ij}^{(r)}\right|\alpha_{ij}^2,
\end{align}
where $u$ is a nonnegative constant, is satisfied, the path $\{x_{ij}^{(r)}\}_{i=1,n_c}^{j=1,t}$  survives with label $\calC$.  Otherwise, path $\{x_{ij}^{(r)}\}_{i=1,n_c}^{j=1,t}$ survives with label $\calX$. Besides, if $\{x_{ij}^{(r)}\}_{i=1,n_c}^{j=1,t}$ has label $\calX$ at level $t-1$, it will have the label $\calX$ at level $t$. Retransmission is requested if at level $H+m$ all the survivors are labeled with $\calX$. 

Given the fixed code-length $H$, by making use of the dynamic programming of LHS and RHS of~\eqref{eq7keykey}, the complexity of this overall decoding scheme is $O(2^K H)$ since finding the paths  $\{x_{ij}^{(r)}\}_{i=1,n_c}^{j=1,H+m}$ and $\{(x_{ij}^{(r)})'\}_{i=1,n_c}^{j=1,H+m}$, which have the largest sum of branch metrics $\sum_{j=1}^{H+m}\lambda_j^{(r)}$ and the second largest one as in the Yamamoto-Itoh algorithm (or in the Viterbi algorithm), requires $O(2^K H)$ computations~\cite{Rimoldibook}. This means that the complexity of this proposed decoding algorithm is exponential in the convolutional code constraint length and linear in the convolutional code length. Note that for the polar code~\cite{Arikan}, the complexity is $O(H\ln H)$. However, it should be mentioned that the constraint length $K$ has an effect on the reliability of the convolutional code (or free distance of the code $d_{f})$. When $u=0$, this decoding strategy coincides with the traditional MLSD (Maximum Likehood Sequence Decoding) decoding scheme (or the original Viterbi decoding for the no-ARQ case). 
\begin{remark} \label{rmk1} \normalfont Some remarks about the modified Yamamoto-Itoh algorithm above are given below.
\begin{itemize}
\item The constraint~\eqref{eq7keykey} is slightly different from the original Yamamoto-Itoh constraint in~\cite[Eq.~(30)]{YamamotoItoh1980} to avoid the appearance of the inverse of fading terms in the error probability when $u>0$ if directly using the Yamamoto-Itoh algorithm. This modification is very useful in fading channels where the Rician factor, $\gamma$, is very small. It makes all terms inside integrals to be finite even if channel fading coefficients approach zero. For $u=0$, this modified algorithm coincides with the Viterbi decoding algorithm~\cite[Eq.~(22)]{Viterbi1971}.
\item There are some improved algorithms (with higher complexity) of the Yamamoto-Itoh algorithm,~\cite{YamamotoItoh1979}, with respect to the error exponent,~\cite{Raghavan1998},~\cite{Hashimoto1999}, however for the purpose of analyzing performance (bit error probability and retransmission probability as functions of $E_c/N_0$ and $\gamma$) in this paper, it is hard to know whether these algorithms are better than the Yamamoto-Itoh algorithm or not. For example, if we assume that $BER=\exp(-\alpha_1 H) (E_c/N_0)^{-\alpha_2} \exp(-\gamma \alpha_3)$, then the improvement in the error exponent with respect to block-length $H$, i.e., $\alpha_1$, does not mean the improvement over the decay rate of error probability with respect to $E_c/N_0$, i.e., $\alpha_2$, or the improvement over the bit error exponent with respect to Rician factor, i.e., $\alpha_3$ \footnote{Refer to Definition~\ref{def2new} for a formal definition of the bit error exponent with respect to Rician factor.}. In addition, we choose the Yamamoto-Itoh algorithm to analyze performance for the sake of simplicity of mathematical analysis. Furthermore, our results obviously show that at finite code-length $H$, higher complexity decoding schemes such as ROVA~\cite{Raghavan1998},~\cite{Hashimoto1999} can at least achieve the same performance as the Yamamoto-Itoh decoding scheme considered in this paper. However, whether the ROVA~\cite{Raghavan1998},~\cite{Hashimoto1999} and other improved algorithms can strictly improve the error exponent with respect to the Rician factor $\gamma$ is still open.
\end{itemize}
\end{remark}
\section{Some Mathematical Preliminaries}\label{math_pre}
In this paper, we use the notation $x_+=\max\{x,0\}$, and $\bbR_{+}$ is the set of positive real numbers. We also use asymptotic notation such as $O(\cdot)$ and $\Theta(\cdot)$ for showing complexity order; $\tilf(x) = O(\tilg(x))$ holds if and only if $\limsup_{x\to \infty} \tilf(x)/\tilg(x) < C$ for some positive constant $C< \infty$. For the notation $\Theta(\cdot)$, the inequality is replaced by the equality. $\lfloor x \rfloor$ is the standard floor function. The indicator function $\bone\{A\}=1$ if $A$ is true, and $\bone\{A\}=0$ otherwise. The average pairwise error probability per branch (or the first-event error probability~\cite{Viterbi1971}), the average bit error probability, and the average retransmission probability of the Yamamoto-Itoh algorithm with the Yamamoto-Itoh flag $u$,\cite{YamamotoItoh1980}, are denoted by $\rvP_{\rme}(u),\rvP_{\rmb}(u),\rvP_{\rmx}(u)$. For $u=0$, the Yamamoto-Itoh algorithm is the original Viterbi decoding algorithm, so $\rvP_{\rme}(0)$ and $\rvP_{\rmb}(0)$ are the average pairwise error probability and average bit error probability of the original Viterbi decoding,\cite{Viterbi1967},~\cite{Viterbi1971}, respectively. Note that $\rvP_{\rmx}(0)=0$. Given a fading realization $\alpha$, $\rvP_{\rme}(u|\alpha),\rvP_{\rmb}(u|\alpha)$ and $\rvP_{\rmx}(u|\alpha)$ are used to indicate corresponding conditional performance measures.
\begin{definition}\label{def2new} The average bit error exponent of a convolutional code with/without ARQ with respect to Rician factor is defined as
$
\lim_{\gamma\to \infty}-\ln \rvP_{\rmb}(u)/\gamma.
$
\begin{remark}\normalfont The average bit error exponent of a convolutional code with/without ARQ with respect to Rician factor in Definition~\ref{def2new} is different from the traditional error exponent with respect to code block-length which is defined as
$
\lim_{H\to \infty}-\ln \rvP_{\rmb}(u)/H.
$
\end{remark}
\end{definition}
Next, we prove two preliminary lemmas which will be used later in upper bounding and approximating reliability performances.
\begin{lemma}
\label{lemeasy}
For any variables $\Phi_1>0,\Phi_2\in \mathbb{R},  z>0$, where $\Phi_1$ and $\Phi_2$ can be dependent on each other but $z$ is independent of $\Phi_1$ and $\Phi_2$\footnote{Here, two variables are said to be dependent if one variable is a function of the other variable. Two variables are said to be independent if no variable is a function of the other.}, define
\begin{align}
\label{definevarphi}
\varphi(\Phi_1,\Phi_2,z)&:=\int_0^{z} \alpha \exp\left(-\Phi_1 \alpha^2 -\Phi_2 \alpha\right) d\alpha.
\end{align}
Then, the following expression holds:
\begin{align}
\label{easyresult2}
&\varphi(\Phi_1,\Phi_2,\infty) \nonumber\\
&\quad =\frac{1}{2\Phi_1}\left[1-\frac{\Phi_2\sqrt{2\pi}}{\sqrt{2\Phi_1}} \exp\left(\frac{\Phi_2^2}{4\Phi_1}\right) Q\left(\frac{\Phi_2}{\sqrt{2\Phi_1}}\right)\right], 
\end{align}
where $Q(x)$ is defined as 
\begin{align}
\label{defineQ}
Q(x):=\frac{1}{\sqrt{2\pi}}\int_x^{\infty} \exp\left(-\frac{t^2}{2}\right) dt, \quad \forall x\in \mathbb{R}.
\end{align}
\end{lemma}
\begin{IEEEproof}
We provide the proof of Lemma~\ref{lemeasy} in Appendix~\ref{app:prf_moments}.
\end{IEEEproof}
\begin{lemma}
\label{lem1}
Consider a convolutional code with transfer function or code power series $T(D,N)$\footnote{The transfer function for convolutional codes is defined by Viterbi~\cite{Viterbi1971}.}. Assume that
\begin{align}
\label{eq14t}
T(D,N)&=\sum_{k=d_f}^{\infty} a_k D^k N^{l_k},\\
\label{eq15t}
\frac{dT(D,N)}{dN}\Big|_{N=1}&=\sum_{k=d_f}^{\infty} c_k D^k,
\end{align}
where $\{a_k\}_{k=1}^{\infty}$ and $\{l_k\}_{k=1}^{\infty}$ are coefficients and exponents of $N$ in the power series expansions of the transfer function, respectively.  In addition, $\{c_k\}_{k=1}^{\infty}$ are coefficients in the the power series expansions of the derivative of the transfer function. Then, the following bounds hold:
\begin{align}
\label{boundak}
a_k&\leq 4^k,\\
\label{boundck}
c_k&\leq k k_c a_k.
\end{align}
\end{lemma}
\begin{IEEEproof}
Refer to Appendix~\ref{app:prf_lem1} for a detailed proof.
\end{IEEEproof}
\begin{remark} \label{rm32018} Some remarks about Lemma~\ref{lem1} are given below.
\begin{itemize}
\item The intermediate $D$ in~\eqref{eq14t} and~\eqref{eq15t} has the meaning of delay, and $D$ is sometimes called the delay operator.
\item The exponents of the factors in $N$ in each term determine the number of bit errors for the path(s) corresponding to that term.
\item The exponents of the factors in $D$ in each term determine the distance of the path(s) from the all zeros path.
\item Upper bounds in~\eqref{boundak} and~\eqref{boundck} can be made even tighter. However,~\eqref{boundak} and~\eqref{boundck} are good enough for our purpose of deriving the asymptotic decay rate of the bit error probability and the average error exponent of convolutional codes. 
\item For better bounds of performance at $H$ finite, the exact values of $a_k$ and $c_k$ in the power series expansion of the transfer function $T(D,N)$ of convolutional codes should be used. Those exact values of $a_k$ and $c_k$ are also used in our numerical evaluations to compare with the Monte-Carlo simulation results in Section~\ref{sec:numerical}.
\end{itemize}
\end{remark}
\section{Main Results} \label{main}
In this section, we state our main contributions in this paper. The decay rate of the average bit error probability with respect to the bit energy to noise ratio is given for both cases (with ARQ and without ARQ). The effects of the Yamamoto-Itoh flag on the average bit error probability are also considered. Our main contribution in this paper is the following theorem. 
\begin{theorem} \label{thm} By using the modified Yamamoto-Itoh decoding algorithm over a Rician flat fading channel with unlimited interleaving in Subsection~\ref{decodalg}, the following hold.
\begin{itemize}
\item The decay rate of the bit error probability $\rvP_{\rmb}(0)$ of the original Viterbi decoding scheme satisfies the following constraint
\begin{align}
\label{main1}
\rvP_{\rmb}(0)  \leq O\left(\left(\frac{E_b}{N_0}\right)^{-d_f}\right)
\end{align} for any pair of fixed channel parameters $\gamma,\sigma$.
\item If the channel allows ARQ, then there exists a Yamamoto-Itoh flag $u_0>0$ such that the retransmission probability $\rvP_{\rmx}(u_0)$ and the decay rate of the bit error probability $\rvP_{\rmb}(u_0)$ for the Yamamoto-Itoh algorithm satisfy
\begin{align}
\label{main2}
\rvP_{\rmx}(u_0)&=O\left(\left(\frac{E_b}{N_0}\right)^{-d_f}\right),\\
\label{main3}
\rvP_{\rmb}(u_0) &\leq 4^{-d_f} \rvP_{\rmb}(0).
\end{align}
\item In addition, for any fixed transmission bit energy to noise ratio $(E_b/N_0)$ and Yamamoto-Itoh flag $u\geq 0$, the following hold:
\begin{align}
\label{eq23best}
\liminf_{\gamma\to \infty} \frac{-\ln \rvP_{\rmb}(u)}{\gamma} &\geq d_f \tilh(u),\\
\label{star10}
\liminf_{\gamma\to \infty} \frac{-\ln \rvP_{\rmx}(u)}{\gamma} &\geq d_f \tilh(-u), \quad u<d_f\sqrt{2E_c/N_0}
\end{align} for any convolutional code using the Yamamoto-Itoh algorithm for decoding, where
\begin{align}
\label{deftilh}
\tilh(u):=1-\frac{1}{\sigma^2}\left[\left(\sqrt{\frac{2E_c}{N_0}}+\frac{u}{d_f}\right)^2+\frac{1}{\sigma^2}\right]^{-1}.
\end{align}
\end{itemize}
\end{theorem}
\begin{IEEEproof}
The proof of Theorem~\ref{thm} is provided in Section~\ref{sec:lower}. 
\end{IEEEproof}
\begin{remark}\label{rmk100}
\normalfont
Some remarks about Theorem~\ref{thm} are given below.
\begin{itemize}
\item The results in~\eqref{main1},~\eqref{main2}, and~\eqref{main3} can be intuitively explained as follows. The Viterbi decoding error is dominated by a term which is proportional to the pairwise error probability between two codewords of the Hamming distance equal to $d_f$. This pairwise error probability is equal to $\bbE\Big[Q\Big[\left(\sqrt{\frac{2E_c}{N_0}}\right)\sqrt{\sum_{r=1}^{d_f}\alpha_r^2}\Big]\Big]$ where $\{\alpha_r\}_{r=1}^{d_f}$ are $d_f$ independent copies of the Rician fading process~\cite[~(23)]{Viterbi1971}. There are $d_f$ independent random variables in the RHS of this formula, so we conjecture that the decay rate of bit error probability with respect to $E_c/N_0$ to be equal to $d_f$ (time diversity). In addition, the Yamamoto-Itoh flag $u$ plays a role as a power control parameter thanks to ARQ. If the Yamamoto-Itoh algorithm helps to reduce the bit error probability from $\rvP_{\rmb}(0)=\beta (E_c/N_0)^{-d_f}$ to $\rvP_{\rmb}(u)=\beta (E_c/N_0+\tilg(u))^{-d_f}$, it also increases the retranmission rate from $\rvP_{\rmx}(0)=\zeta (E_c/N_0)^{-d_f}$ to $\rvP_{\rmx}(u)=\zeta (E_c/N_0-\tilg(u))^{-d_f}$ for some function $\tilg(u): \bbR_{+} \to \bbR_{+}$ and a pair of positive numbers $(\beta, \zeta)$ which does not depend on $E_c/N_0$. For $\rvP_{\rmx}(u)$ being vanished at $E_c/N_0 \to \infty$, we should choose $u_0$ such that $\tilg(u_0) \approx E_c/N_0$. Hence, we conjecture that the optimal Yamamoto-Itoh flag helps to increase the reliability function by a factor of $4^{d_f}$ from this observation.
\item The original Viterbi decoding,\cite{Viterbi1967}, is a special case of Yamamoto-Itoh algorithm when setting the flag $u=0$~\cite{YamamotoItoh1980}. The upper bounds work the same for two cases: with ARQ $(u>0)$ and the original Viterbi decoding~\cite{Viterbi1967}.
\item By setting $\tilde{\alpha}_{ij}=\alpha_{ij}/(\sigma \sqrt{2}), \tilde{s}=s/(\sigma\sqrt{2})$, and $\tilde{E}_c=2\sigma^2 E_c$, from~\eqref{eqn:eqkey1} and~\eqref{eqkey2}, we obtain an equivalent Rician fading channel model with the set of parameter $(\alpha_i, s, E_c, \sigma^2)$ to be replaced by a new set of parameters $(\tilde{\alpha}_i, \tilde{s},\tilde{E}_c, 1/2)$. It is easy to see that this equivalent Rician fading channel model has the same Rician factor $\gamma$ and bit error probability as the setup one in Section~\ref{ch:model}. Furthermore, with this equivalent channel model we have $\tilde{E}_c/N_0=2\sigma^2 E_c/N_0$. Hence, if we fix $\gamma$, $E_c/N_0$, the decay rate of the bit error probability with respect to $\sigma^2$ for the Rician fading channel model in Section~\ref{ch:model} is the same as the decay rate of this bit error probability with respect to $\tilde{E}_c/N_0$ for a fixed $\gamma$. Therefore, we are not interested in finding the effect of $\sigma$ in this paper. This parameter always behaves as the $E_b/N_0$ from this viewpoint.
\item Although the lower bound on the decay rate of the bit error probability with respect to the bit energy per noise ratio for the case $u=0$ (without ARQ) can be found in~\cite[(14.4-40)]{Proakis}, the upper bound versions in~\eqref{main2} and~\eqref{main3} for $u_0>0$ for the Rician channel with ARQ and the effects of Yamamoto-Itoh flags on performance are our new contributions.
\item As $E_c/N_0 \to \infty$, we have $\lim_{\gamma \to \infty}-\ln \rvP_{\rmb}(u)/\gamma=d_f$ for any $u\geq 0$. Hence, the average bit error exponent of a convolutional code with/without ARQ with respect to Rician factor is shown to be positive and approximately equal to $d_f$ at high $E_b/N_0$.
\item Yamamoto and Itoh,\cite{YamamotoItoh1980}, proposed the use of the flag $u$ with the original Viterbi decoding,\cite{Viterbi1967}, for channels which allows ARQ and showed that there exists a value $u_0$ such that the error exponent with respect to code-length $-\lim_{H\to \infty} \ln\rvP_{\rmb}(u_0)/H$ is doubled compared with using the original Viterbi decoding for DM channels without ARQ,~\cite{Viterbi1967}, for DM channels. For the fading channel, this paper shows that this flag can help to reduce the error probability by a factor of $4^{d_f}$ with the same parameters $(E_b/N_0, H,m,d_f,R_c, k_c)$ and negligible retransmission probability at high $E_b/N_0$. This result is useful if $d_f=\Theta(\log(\frac{E_b}{N_0}))$. 
\end{itemize}
\end{remark}
\section{Proof of Theorem~\ref{thm}}\label{sec:lower}
\subsection{General Upper Bounds for finite values of $H$ and $(E_b/N_0)$}
\begin{proposition} \label{prop1} For any convolutional code with transfer function $T(D,N)$ which uses the Yamamoto-Itoh algorithm with flag $u$ for decoding, the following inequalities hold:
\begin{align}
\label{result1}
\rvP_{\rme}(u)&\leq  \sum_{k=d_f}^{n_c(H+m)} a_k \left[\tilD(E_c/N_0, u/d_f,\sigma, s)\right]^k, \\
\label{result2}
\rvP_{\rmx}(u) &\leq \sum_{k=d_f}^{n_c(H+m)} a_k \left[\tilD(E_c/N_0, -u/d_f,\sigma, s)\right]^k \nonumber\\
&\qquad \qquad \qquad \forall u<d_f \sqrt{2E_c/N_0},\\
\label{result3}
\rvP_{\rmb}(u)& \leq \sum_{k=d_f}^{n_c(H+m)}  c_k  \left[\tilD(E_c/N_0, u/d_f,\sigma, s)\right]^k,
\end{align}
where  
\begin{align}
\label{eq27key}
\tilD(E_c/N_0, u/d_f,\sigma, s):=\frac{1}{\pi} \exp\left(-\gamma\right) \int_{0}^{\pi} \Lambda(\theta)d\theta,
\end{align} and
\begin{align}
\label{defineA}
A&:= \frac{1}{2}\left(\sqrt{\frac{2E_c}{N_0}}+\frac{u}{d_f}\right)^2+\frac{1}{2\sigma^2},\\
\label{defineB}
B_{\theta}&:=-\frac{s\cos \theta}{\sigma^2},\\
\label{defLambda}
\Lambda(\theta)&:=\frac{1}{2A\sigma^2}\left[1-\frac{B_{\theta}\sqrt{2\pi}}{\sqrt{2A}}\exp\left(\frac{B_{\theta}^2}{4A}\right)Q\left(\frac{B_{\theta}}{\sqrt{2A}}\right)\right].
\end{align}
Here, $\{a_k\}_{k=1}^{\infty}$ and $\{c_k\}_{k=1}^{\infty}$ are defined in~\eqref{eq14t} and~\eqref{eq15t}.
\end{proposition}
\begin{corollary}\label{corbet}
 For any convolutional code with transfer function $T(D,N)$ which uses the Yamamoto-Itoh algorithm with flag $u$ for decoding, the following inequalities hold:
\begin{align}
\label{eq32h}
\rvP_{\rme}(u) & \leq T(D,N)\Big|_{N=1\atop D=\tilD(E_c/N_0, u/d_f,\sigma^2, s)},\\
\rvP_{\rmx}(u)&\leq  T(D,N)\Big|_{N=1\atop D=\tilD(E_c/N_0, -u/d_f,\sigma^2, s)} \nonumber \\
&\quad \quad \quad \forall u < d_f \sqrt{\left(\frac{2E_b}{N_0}\right)\left(\frac{H}{H+m}\right)R_c}, \\
\label{eq34h}
\rvP_{\rmb}(u)& \leq \frac{dT(D,N)}{dN}\Big|_{N=1\atop D=\tilD(E_c/N_0, u/d_f,\sigma^2, s)}.
\end{align}
\end{corollary}
\begin{IEEEproof}[Proof of Proposition~\ref{prop1} and Corollary~\ref{corbet}]
Assume that $\{x_{ij}^{(r)}\}_{i=1,n_c}^{j=1,H+m}$ is a transmitted coded sequence. Since we assume perfect CSI at the receiver, from~\eqref{eq7keykey}, the error event is 
\begin{align}
\label{eq24keynew}
\{V_t \geq W_t\}
\end{align} for some $t\in \{1,2,\ldots,H+m\}$, where
\begin{align}
V_t&:=\sum_{j=1}^t \sum_{i=1}^{n_c} \alpha_{ij} \left((x_{ij}^{(r)})'- x_{ij}^{(r)}\right) y_{ij},\\
W_t&:=\frac{u}{d_f}  \sqrt{N_0/2}\sum_{j=1}^t \sum_{i=1}^{n_c}\left|(x_{ij}^{(r)})'- x_{ij}^{(r)}\right|\alpha_{ij}^2.
\end{align} Here, $\{(x_{ij}^{(r)})'\}_{i=1,n_c}^{j=1,t}$ is another path which merges with the transmitted sequence $\{x_{ij}^{(r)}\}_{i=1,n_c}^{j=1,t}$ at time $t$. In addition, if one of the conditions~\eqref{eq7keykey} (correct event) or~\eqref{eq24keynew} (error event) happens, the modified Yamamoto-Itoh algorithm put the label $\calC$ on the survivor at this merging time $t$, which does not cause the retransmission to happen. Hence, the retransmission event at each merging time $t$ of the two paths is a subset of the event
\begin{align}
\label{eq24rx}
\{W_t > V_t > -W_t\}. 
\end{align}
Assume that at the merging time $t$ the Hamming distance between the two sequences $\{x_{ij}^{(r)}\}_{i=1,n_c}^{j=1,t}$ and $\{(x_{ij}^{(r)})'\}_{i=1,n_c}^{j=1,t}$ is $k \in \{d_f,d_f+1,\ldots,H+m\}$. It follows that the conditional pairwise error probability at the merging time $t$ which is caused by incorrectly choosing the survivor with $\calC$ at the merging node is bounded by
\begin{align}
\label{def35pku}
&\rvP^{(t)}_{\rmk,\rme}(u|\alpha):=\bbP\big(V_t \geq W_t \big)\\
\label{41eq}
&\quad=\bbP\left(\sum_{r=1}^k  \alpha_r y_r \leq  -\frac{u }{d_f}  \sqrt{N_0/2}\sum_{r=1}^{k}\alpha_r^2  \right).
\end{align}
Here,~\eqref{41eq} follows from~\cite{Viterbi1971}, $y_r \sim \calN(\alpha_r \sqrt{E_c}, N_0/2)$, and $\alpha_1,\alpha_2,\ldots,\alpha_{n_c(H+m)}$ is a permutation of $\{\alpha_{ij}\}_{i=1,n_c}^{j=1,H+m}$. 

Denote by
\begin{align}
\label{defqku}
q_k(u):=Q\left[\left(\sqrt{\frac{2E_c}{N_0}}+\frac{u}{d_f}\right)\sqrt{\sum_{r=1}^{k}\alpha_r^2}\right],
\end{align} where the function $Q(x)$ is defined in~\eqref{defineQ}.

Now, since
\begin{align}
\label{42eq}
\sum_{r=1}^k \alpha_r y_r \sim \mathcal{N}\left(\sum_{r=1}^k \sqrt{E_c} \alpha_r^2, \frac{N_0}{2}\sum_{r=1}^k \alpha_r^2\right),
\end{align}
the bound in~\eqref{41eq} does not depend on $t$. In addition, from~\eqref{41eq} and~\eqref{42eq}, the conditional pairwise error probability which is caused by choosing an incorrect survivor with Hamming distance $k$ from the correct one is 
\begin{align}
\label{pkerror}
\rvP_{\rmk,\rme}(u|\alpha) =q_k(u),
\end{align} where $q_k(u)$ is defined in~\eqref{defqku}.

Similarly, from~\eqref{eq24rx}, the conditional probability that a path is surviving with the label $\calX$ at the merging time $t$ is
\begin{align}
\rvP^{(t)}_{\rmk,\rmx}(u|\alpha)&:= \bbP\big(\big|V_t\big| < W_t\big)\\
&=\bbP\big(V_t \geq -W_t\big) -\bbP\big(V_t\geq W_t\big)\\
\label{eqpkerror1}
&=q_k(-u)-q_k(u),
\end{align}
where~\eqref{eqpkerror1} follows from~\eqref{def35pku} and~\eqref{pkerror}. Since the bound in~\eqref{eqpkerror1} does not depend on $t$, it follows from~\eqref{eqpkerror1} that the conditional retransmission probability caused by choosing neither the correct survivor nor an incorrect survivor with Hamming distance $k$ from the correct one, $\rvP_{\rmk,\rmx}(u|\alpha)$, satisfies
\begin{align}
\label{tempre}
\rvP_{\rmk,\rmx}(u|\alpha)= q_k(-u)-q_k(u).
\end{align}

For the case $u=0$ (Viterbi decoding without Yamamoto-Itoh flag), this probability is equal to zero. This means that
\begin{align}
\rvP_{\rmk,\rmx}(0)=\bbE\left[\rvP_{\rmk,\rmx}(0|\alpha)\right]=0.
\end{align}
or no-retransmission in this coding scheme as expected. 

Now, since $Q(x) \geq 0, \enspace \forall x$, for any $u>0$, we have from~\eqref{tempre} that
\begin{align}
\label{pkretransmision}
\rvP_{\rmk,\rmx}(u|\alpha) \leq q_k(-u).
\end{align}

Hence, for a given realization of fast fading, i.e. $\alpha=(\alpha_1,\alpha_2,\alpha_3, \ldots)$, by Viterbi~\cite{Viterbi1971},~\eqref{pkerror}, and~\eqref{pkretransmision}, the performance functions satisfy: 
\begin{align}
\label{neweq11}
\rvP_{\rme}(u|\alpha) & \leq \sum_{k=d_f}^{n_c(H+m)}a_k   q_k(u),\\
\rvP_{\rmx}(u|\alpha) &\leq \sum_{k=d_f}^{n_c(H+m)}a_k  q_k(-u),\\
\label{neweq13}
\rvP_{\rmb}(u|\alpha) &\leq \sum_{k=d_f}^{n_c(H+m)}c_k q_k(u),
\end{align}
where $\rvP_{\rmb}(u|\alpha)$ is the conditional bit error probability caused by choosing an incorrect survivor with Hamming distance $k$ from the correct one.

Now, define
\begin{align}
\label{deflalphar}
\lambda(u,\alpha_r):=\exp\bigg(-\frac{1}{2}\left(\sqrt{\frac{2E_c}{N_0}}+\frac{u}{d_f}\right)^2\alpha_r^2\bigg).
\end{align}
By the fact that $Q(x) \leq \exp(-x^2/2)$ for all $x\geq 0$, we have from~\eqref{defqku} and~\eqref{deflalphar} that
\begin{align}
\label{porcupinefact}
q_k(u) \leq \prod_{r=1}^k \lambda(u,\alpha_r).
\end{align}

Therefore, we have
\begin{align}
\label{eq46final}
\rvP_{\rme}(u)&= \int_0^{\infty} \int_0^{\infty} \cdots \int_0^{\infty} \rvP_{\rme}(u|\alpha) \nonumber\\
&\qquad \times \prod_{k=1}^{n_c(H+m)} f(\alpha_k) d\alpha_1 d\alpha_2\cdots d\alpha_{n_c(H+m)}\\
\label{eq47final}
&\leq  \sum_{k=d_f}^{n_c(H+m)} a_k  \int_0^{\infty} \cdots \int_0^{\infty}    q_k(u)\nonumber\\
&\qquad \times \prod_{r=1}^k f(\alpha_r) d\alpha_1 d\alpha_2\cdots d\alpha_k\\
\label{eq48fin}
&\leq \sum_{k=d_f}^{n_c(H+m)} a_k  \int_0^{\infty} \cdots \int_0^{\infty} \prod_{r=1}^k \lambda(u,\alpha_r) \nonumber\\
&\qquad \times \prod_{r=1}^k f(\alpha_r) d\alpha_1 d\alpha_2\cdots d\alpha_k\\
\label{eq26}
&= \sum_{k=d_f}^{n_c(H+m)} a_k \bigg[\int_0^{\infty} \lambda(u,\alpha_1) f(\alpha_1) d\alpha_1\bigg]^k,
\end{align} where~\eqref{eq46final} follows from~\eqref{eqkey2},~\eqref{eq47final} follows from~\eqref{neweq11} and the fact that $q_k(u)$ is a function of $(\alpha_1,\alpha_2,\cdots,\alpha_k)$ for each fixed $u$,~\eqref{eq48fin} follows from~\eqref{porcupinefact}, and~\eqref{eq26} follows from Fubini's theorem~\cite{Billingsley} and the fact that $\alpha_1,\alpha_2,\ldots, \alpha_{n_c(H+m)}$ are i.i.d. \\
Observe that
\begin{align}
&\int_0^{\infty} \lambda(u,\alpha_1) f(\alpha_1) d\alpha_1=\int_0^{\infty} \lambda(u,\alpha_1) \frac{\alpha_1}{\sigma^2}  \nonumber\\
&\qquad \qquad \times e^{-((\alpha_1^2+s^2)/2\sigma^2)}\frac{1}{2\pi}\int_{-\pi}^{\pi}\exp\left(\frac{\alpha_1 s}{\sigma^2} \cos \theta \right)d\theta d\alpha_1 \\
\label{eq27}
&\quad =\frac{1}{2\pi} \exp\left(-\frac{s^2}{2\sigma^2}\right)\int_{-\pi}^{\pi}\int_0^{\infty}\lambda(u,\alpha_1) \frac{\alpha_1}{\sigma^2}  \nonumber\\
&\qquad \qquad \times \exp\left(-\frac{\alpha_1^2}{2\sigma^2}\right) \exp\left(\frac{\alpha_1 s}{\sigma^2} \cos \theta \right) d\alpha_1d\theta.
\end{align}

Now, from Lemma~\ref{lemeasy} we have
\begin{align}
\label{eq28newnew}
&\int_0^{\infty}\lambda(u,\alpha_1) \frac{\alpha_1}{\sigma^2} \exp\left(-\frac{\alpha_1^2}{2\sigma^2}\right) \exp\left(\frac{\alpha_1 s}{\sigma^2} \cos \theta \right) d\alpha_1\nonumber\\ 
&=\frac{1}{\sigma^2} \varphi\left( \frac{1}{2}\left(\sqrt{\frac{2E_c}{N_0}}+\frac{u}{d_f}\right)^2+\frac{1}{2\sigma^2}, -\frac{s\cos\theta}{\sigma^2}, \infty  \right)\\
\label{eq27newkey}
&=\Lambda(\theta).
\end{align} 
Hence, we obtain
\begin{align}
&\int_0^{\infty}\lambda(u,\alpha_1) \frac{\alpha_1}{\sigma^2} \exp\left(-\frac{\alpha_1^2}{2\sigma^2}\right) \exp\left(\frac{\alpha_1 s}{\sigma^2} \cos \theta \right) d\alpha_1\nonumber\\
\label{stack16}
&=\frac{1}{\pi} \exp\left(-\gamma\right) \int_{0}^{\pi} \Lambda(\theta)d\theta\\
\label{stack17}
&=\tilD(E_c/N_0, u/d_f,\sigma, s).
\end{align} Here,~\eqref{stack16} follows from the fact that $\Lambda(\theta)$  is an even function in $\theta$ since $B_{\theta}=-s\cos \theta/\sigma^2$ is even in $\theta$ and the fact that $\gamma=s^2/(2\sigma^2)$, and~\eqref{stack17} follows from~\eqref{eq27key}.\\
From~\eqref{eq26},~\eqref{eq27}, and~\eqref{eq27newkey}, we obtain~\eqref{result1} in Proposition~\ref{prop1}. Similarly, we obtain~\eqref{result2} and~\eqref{result3} of Proposition~\ref{prop1}. The further upper bounds in Corollary~\ref{corbet} can be obtained by taking the limit $H\to \infty$ of~\eqref{result1},~\eqref{result2} and~\eqref{result3}, and using~\eqref{eqeb}. This concludes our proof of Proposition~\ref{prop1} and Corollary~\ref{corbet}. 
\end{IEEEproof}
\subsection{Reliability Evaluations}
\begin{IEEEproof}[Proof of Theorem~\ref{thm}] Observe from~\eqref{eq27key} that
\begin{align}
&\tilD(E_c/N_0, u/d_f,\sigma, s)\leq \frac{1}{\pi} \exp\left(-\gamma\right) \int_{0}^{\pi/2} \Lambda(\theta)d\theta \nonumber\\
\label{eq84b}
&\qquad + \frac{1}{\pi} \exp\left(-\gamma\right) \int_{\pi/2}^{\pi} \frac{1}{2A\sigma^2} d\theta\\
\label{eq71:final}
&\leq \frac{1}{\pi} \exp\left(-\gamma\right) \int_{0}^{\pi/2} \Lambda(\theta)d\theta + \frac{\exp(-\gamma)}{4A\sigma^2},
\end{align} where~\eqref{eq84b} follows from the fact that $B_{\theta} \geq 0$ (so $\Lambda(\theta)\leq 1/(2A\sigma^2)$ by~\eqref{defLambda}) for all $\pi/2\leq \theta \leq \pi$.

Now, we note that
\begin{align}
&\sigma^2 \int_{0}^{\pi/2} \Lambda(\theta) d\theta\nonumber\\
\label{eq72:final}
&=\frac{\pi}{4A}-\frac{\sqrt{2\pi}}{2A \sqrt{2A}}\int_0^{\pi/2} B_{\theta} \exp\left(\frac{B_{\theta}^2}{4A}\right)Q\left(\frac{B_{\theta}}{\sqrt{2A}}\right)d\theta.
\end{align}
It follows from~\eqref{eq71:final} and~\eqref{eq72:final} that
\begin{align}
\label{deftilDapp}
&\tilD(E_c/N_0, u/d_f,\sigma, s)\leq \frac{1}{2A\sigma^2}\exp(-\gamma)-\frac{1}{\pi} \exp\left(-\gamma\right)\nonumber\\
&\quad \times \left[\frac{\sqrt{2\pi}}{2A \sigma^2 \sqrt{2A}}\int_0^{\pi/2} B_{\theta} \exp\left(\frac{B_{\theta}^2}{4A}\right)Q\left(\frac{B_{\theta}}{\sqrt{2A}}\right)d\theta\right].
\end{align}
Now, we consider two cases:
\begin{itemize}
\item \emph{Case 1: $\gamma$ and $\sigma$ are fixed.}
\end{itemize}
For this case, observe that
\begin{align}
\chi(A)&:=-\exp\left(-\gamma\right)\int_0^{\pi/2} B_{\theta} \exp\left(\frac{B_{\theta}^2}{4A}\right)Q\left(\frac{B_{\theta}}{\sqrt{2A}}\right)d\theta\\
\label{eq91b}
&\leq -\exp\left(-\gamma\right)\int_0^{\pi/2} B_{\theta} \exp\left(\frac{B_{\theta}^2}{4A}\right)d \theta\\
\label{eq91c}
&\leq \exp\left(-\gamma\right)\int_0^{\pi/2} \frac{s}{\sigma^2} \exp\left(\frac{s^2}{4A\sigma^4}\right)d \theta,\\
\label{applex}
&=\frac{\pi}{2} \exp\left(-\gamma\right) \frac{s}{\sigma^2} \exp\left(\frac{s^2}{4A\sigma^4}\right).
\end{align}
Here,~\eqref{eq91b} follows from the fact that $B_{\theta}=-s\cos\theta/\sigma^2\leq 0$ for all $0\leq \theta\leq \pi/2$ which leads to the term inside the integral to be non-positive and the fact that $Q(x)\leq 1,\forall x\in \bbR$,~\eqref{eq91c} follows from the fact that $|B_{\theta}|=s |\cos \theta|/\sigma^2 \leq s/\sigma^2$. 

It follows from~\eqref{deftilDapp} and~\eqref{applex} that
\begin{align}
&\tilD(E_c/N_0, u/d_f,\sigma, s)\leq \frac{1}{2A\sigma^2}\exp(-\gamma)\nonumber\\
&\quad +\frac{1}{\pi}\frac{\sqrt{2\pi}}{2A\sigma^2 \sqrt{2A}}\frac{\pi}{2} \exp\left(-\gamma\right) \frac{s}{\sigma^2} \exp\left(\frac{s^2}{4A\sigma^4}\right)\\
\label{eq:95a}
&=\exp(-\gamma)\left[\frac{1}{2A \sigma^2}+\frac{\sqrt{4\pi \gamma}}{4A \sigma^3\sqrt{2A}}\exp\left(\frac{\gamma}{2A\sigma^2}\right)\right]\\
\label{eq96a}
&=O\left(\frac{1}{A}\right)\quad \mbox{as}\quad A\to \infty.
\end{align}
Therefore, by Lemma~\ref{lem1} and Corollary~\ref{corbet}, we have
\begin{align}
\rvP_{\rmx}(u)&\leq \sum_{k=d_f}^{\infty} a_k [\tilD(E_c/N_0, -u/d_f,\sigma, s)]^k\\
\label{eq100aa}
&\leq \sum_{k=d_f}^{\infty} 4^k [\tilD(E_c/N_0, -u/d_f,\sigma, s)]^k \\
\label{eq100b}
&=\frac{4^{d_f} [\tilD(E_c/N_0, -u/d_f,\sigma, s)]^{d_f}}{1-4 \tilD(E_c/N_0, -u/d_f,\sigma, s)}, \\
\label{eq97newkey}
&=O\left(\left(\sqrt{\frac{2E_c}{N_0}}-\frac{u}{d_f}\right)^{-2d_f}\right)
\end{align}
as $\sqrt{2E_c/N_0}-u/d_f \to \infty$. Here,~\eqref{eq100aa} and~\eqref{eq100b} hold if
\begin{align}
&4 \tilD(E_c/N_0, -u/d_f,\sigma, s) \nonumber\\
&\leq 4\exp(-\gamma)\left[\frac{1}{2A \sigma^2}+\frac{\sqrt{4\pi \gamma}}{4A \sigma^3\sqrt{2A}}\exp\left(\frac{\gamma}{2A\sigma^2}\right)\right]<1
\end{align}
(i.e. $A$ is large enough), and~\eqref{eq97newkey} follows from~\eqref{eq96a} and~\eqref{defineA} where $u$ is replaced by $-u$.

Now, if we  choose $u=u_0$ where
\begin{align}
\label{eq105a}
u_0&:=d_f(1-\delta)\sqrt{\left(\frac{2E_b}{N_0}\right)\left(\frac{H}{H+m}\right)R_c}\\
&=d_f(1-\delta)\sqrt{\frac{2E_c}{N_0}}.
\end{align} for some $\delta>0$. Then, from~\eqref{eq97newkey} and~\eqref{eq105a}, we have
\begin{align}
\label{Rx}
&\rvP_{\rmx}(u_0) =O\left( \left(\frac{E_b}{N_0}\left(\frac{H}{H+m}\right)R_c\right)^{-d_f}\right),
\end{align} so $\rvP_{\rmx}(u_0)\to 0$ as $(E_b/N_0) \to \infty$. 

With this choice of $u_0$, by Proposition~\ref{prop1}, we also have
\begin{align}
\rvP_{\rmb}(u_0)& \leq \sum_{k=d_f}^{n_c(H+m)}  c_k [\tilD(E_c/N_0, u_0/d_f,\sigma, s)]^k\\
\label{eq108a}
&\leq \sum_{k=d_f}^{n_c(H+m)}  k a_k k_c  [\tilD(E_c/N_0, u_0/d_f,\sigma, s)]^k \\
\label{eq190a}
&\leq k_c\sum_{k=d_f}^{n_c(H+m)}  k 4^k  [\tilD(E_c/N_0, u_0/d_f,\sigma, s)]^k \\
\label{eq111a}
&\leq k_c\sum_{k=d_f}^{n_c(H+m)} 6^k  [\tilD(E_c/N_0, u_0/d_f,\sigma, s)]^k \\
\label{eqnear}
&= k_c \frac{6^{d_f} [\tilD(E_c/N_0, u_0/d_f,\sigma, s)]^{d_f}}{1-6 \tilD(E_c/N_0, u_0/d_f,\sigma, s)}\\
\label{eqyamamoto}
&= (2-\delta)^{-2d_f} O\left(\left(\frac{E_b}{N_0}\left(\frac{H}{H+m}\right) R_c\right)^{-d_f}\right),
\end{align}
where~\eqref{eq108a} and~\eqref{eq190a} follow from Lemma~\ref{lem1},~\eqref{eq111a} follows from the fact that $k 4^k \leq 6^k$ for all $k \geq 1$, and~\eqref{eqnear} holds if
\begin{align}
\label{keydoc}
&6 \tilD(E_c/N_0, u/d_f,\sigma, s) \nonumber\\
&=6 \exp(-\gamma)\left[\frac{1}{2A \sigma^2}+\frac{\sqrt{4\pi \gamma}}{4A \sigma^3\sqrt{2A}}\exp\left(\frac{\gamma}{2A\sigma^2}\right)\right]<1,
\end{align}
(i.e. $A$ sufficiently large but finite and the threshold can be estimated even tighter),~\eqref{eqyamamoto} follows from~\eqref{eqeb},~\eqref{defineA},~\eqref{eq96a}, and~\eqref{eq105a}. Hence, we obtain for a pair of fixed channel parameters $(\gamma,\sigma^2)$ that
\begin{align}
\rvP_{\rmb}(u_0) \leq O\left(\left(\frac{E_b}{N_0}\right)^{-d_f}\right),
\end{align}
and
\begin{align}
\rvP_{\rmx}(u_0)=O\left(\left(\frac{E_b}{N_0}\right)^{-d_f}\right).
\end{align}

In addition, by noting that the Viterbi decoding without Yamamoto-Itoh flag corresponds to the case $\delta=1$, from~\eqref{eqyamamoto} and the fact that the bound in~\eqref{result1} is tight for $E_b/N_0$ sufficiently large (cf.~Fig.~\ref{fig2b}), we see that
\begin{align}
\rvP_{\rmb}(u_0) \approx (2-\delta)^{-2d_f} \rvP_{\rmb}(0).
\end{align}
Since $\delta>0$ can be arbitrarily chosen, we have
\begin{align}
\rvP_{\rmb}(u_0) \approx 4^{-d_f} \rvP_{\rmb}(0).
\end{align}
\begin{itemize}
\item \emph{Case 2: $A$ and $\sigma$ are fixed.}
\end{itemize}
For this case, we consider the following function:
\begin{align}
&\tilde{\chi}(\gamma):=-\exp\left(-\gamma\right)\int_0^{\pi/2} B_{\theta} \exp\left(\frac{B_{\theta}^2}{4A}\right)Q\left(\frac{B_{\theta}}{\sqrt{2A}}\right)d\theta\\
&=\int_0^{\pi/2} \exp\left(-\gamma\right) \frac{s\cos\theta}{\sigma^2}  \exp\left(\frac{s^2 \cos^2 \theta}{4A \sigma^4}\right)Q\left(\frac{-s\cos \theta}{\sigma^2 \sqrt{2A}}\right)d\theta\\
&=\int_0^{\pi/2} \exp\left(-\gamma\right) \frac{\sqrt{2\gamma} \cos\theta}{\sigma}  \nonumber\\
&\qquad \times \exp\left(\frac{\gamma \cos^2 \theta}{2A \sigma^2}\right)Q\left(\frac{-\sqrt{2\gamma}\cos \theta}{\sigma \sqrt{2A}}\right)d\theta\\
\label{star1}
&=\sqrt{2\gamma}\exp\left(-\gamma\left(1-\frac{1}{2A\sigma^2}\right) \right) \nonumber\\
&\quad \times \int_0^{\pi/2}  \frac{ \cos\theta}{\sigma}  \exp\left(\frac{-\gamma \sin^2 \theta}{2A \sigma^2}\right)Q\left(\frac{-\sqrt{2\gamma}\cos \theta}{\sigma \sqrt{2A}}\right)d\theta.
\end{align}
Now, define
\begin{align}
\label{star2}
\rho(\gamma,\theta):= \frac{ \cos\theta}{\sigma}  \exp\left(\frac{-\gamma \sin^2 \theta}{2A \sigma^2}\right)Q\left(\frac{-\sqrt{2\gamma}\cos \theta}{\sigma \sqrt{2A}}\right)d\theta.
\end{align}
It is easy to see that $\rho(\gamma,\theta)$ is non-negative and non-increasing for $\theta \in [0,\pi/2]$, integrable from $0$ to $\pi/2$ at $\gamma=0$, and
\begin{align}
\lim_{\gamma \to \infty} \rho(\gamma,\theta)=0. 
\end{align}
Hence, by Dominated Convergence Theorem,\cite{Billingsley}, we have
\begin{align}
&\lim_{\gamma \to \infty} \int_0^{\pi/2}  \frac{ \cos\theta}{\sigma}  \exp\left(\frac{-\gamma \sin^2 \theta}{2A \sigma^2}\right)Q\left(\frac{-\sqrt{2\gamma}\cos \theta}{\sigma \sqrt{2A}}\right)d\theta \nonumber\\
\label{star3}
&=\int_0^{\pi/2}\lim_{\gamma \to \infty} \rho(\gamma,\theta)d\theta =0.
\end{align}
From~\eqref{star1},~\eqref{star2}, and~\eqref{star3}, we obtain
\begin{align}
\tilde{\chi}(\gamma)=\sqrt{2\gamma}\exp\left(-\gamma\left(1-\frac{1}{2A\sigma^2}\right) \right)o(1),
\end{align} as $\gamma \to \infty$.
It follows that for $\gamma$ sufficiently large we have
\begin{align}
&\tilD(E_c/N_0, u/d_f,\sigma, s)\leq  \frac{1}{2A\sigma^2}\exp(-\gamma)\nonumber\\
&\quad+\sqrt{2\gamma}\exp\left(-\gamma\left(1-\frac{1}{2A\sigma^2}\right) \right)o(1),\\
&\leq \frac{1}{2A\sigma^2}\sqrt{2\gamma}\nonumber\\
&\quad \times \exp\left(-\gamma\left(1-\frac{1}{2A\sigma^2}\right)\right)\left[\exp\left(-\frac{\gamma}{2A\sigma^2}\right) +o(1)\right],\\
&\leq \frac{2}{\sigma^2}\sqrt{\frac{\gamma}{2}} \left[\left(\sqrt{\frac{2E_c}{N_0}}+\frac{u}{d_f}\right)^2+\frac{1}{\sigma^2}\right]^{-1}\exp\big(-\gamma\tilh(u)\big)\\
&=O\bigg(\sqrt{\frac{\gamma}{2}}\exp\big(-\gamma\tilh(u)\big)\bigg),
\end{align} as $\gamma \to \infty$.

Now, under the condition $\tilh(u)>0$, we have 
\begin{align}
\rvP_{\rmb}(u)& \leq \sum_{k=d_f}^{n_c(H+m)}  c_k \left[\tilD(E_c/N_0, u/d_f,\sigma^2, s)\right]^k\\
\label{eq:116mod}
&\leq \sum_{k=d_f}^{n_c(H+m)}  k a_k k_c  \left[\tilD(E_c/N_0, u/d_f,\sigma^2, s)\right]^k \\
\label{eq:117mod}
&\leq k_c \sum_{k=d_f}^{n_c(H+m)}  k 4^k \left[\tilD(E_c/N_0, u/d_f,\sigma^2, s)\right]^k\\
&\leq k_c \sum_{k=d_f}^{\infty}  8^k  \left[\tilD(E_c/N_0, u/d_f,\sigma^2, s)\right]^k\\
\label{eq2017key2}
&=k_c \frac{8^{d_f} [\tilD(E_c/N_0, u/d_f,\sigma^2, s)]^{d_f}}{1-8 \tilD(E_c/N_0, u/d_f,\sigma^2, s)}
\end{align} for $\gamma$ sufficiently large, where~\eqref{eq:116mod} and~\eqref{eq:117mod} follow from Lemma~\ref{lem1}.\\
Since $\tilh(u)>0$ for any $u\geq 0$, hence it follows that~\eqref{eq23best} holds. 
Similarly, from~\eqref{result2} and by replacing $u$ with $-u$, we can show that~\eqref{star10} holds under the conditions that $\tilh(-u)>0$ or $u < d_f\sqrt{2E_c/N_0}$. 
This concludes our proof of Theorem~\ref{thm}.
\end{IEEEproof}
\section{Numerical Simulations} \label{sec:numerical}
\begin{figure}[h] 
  \centering
  \begin{minipage}[t]{0.48\textwidth}
    \includegraphics[width=\textwidth]{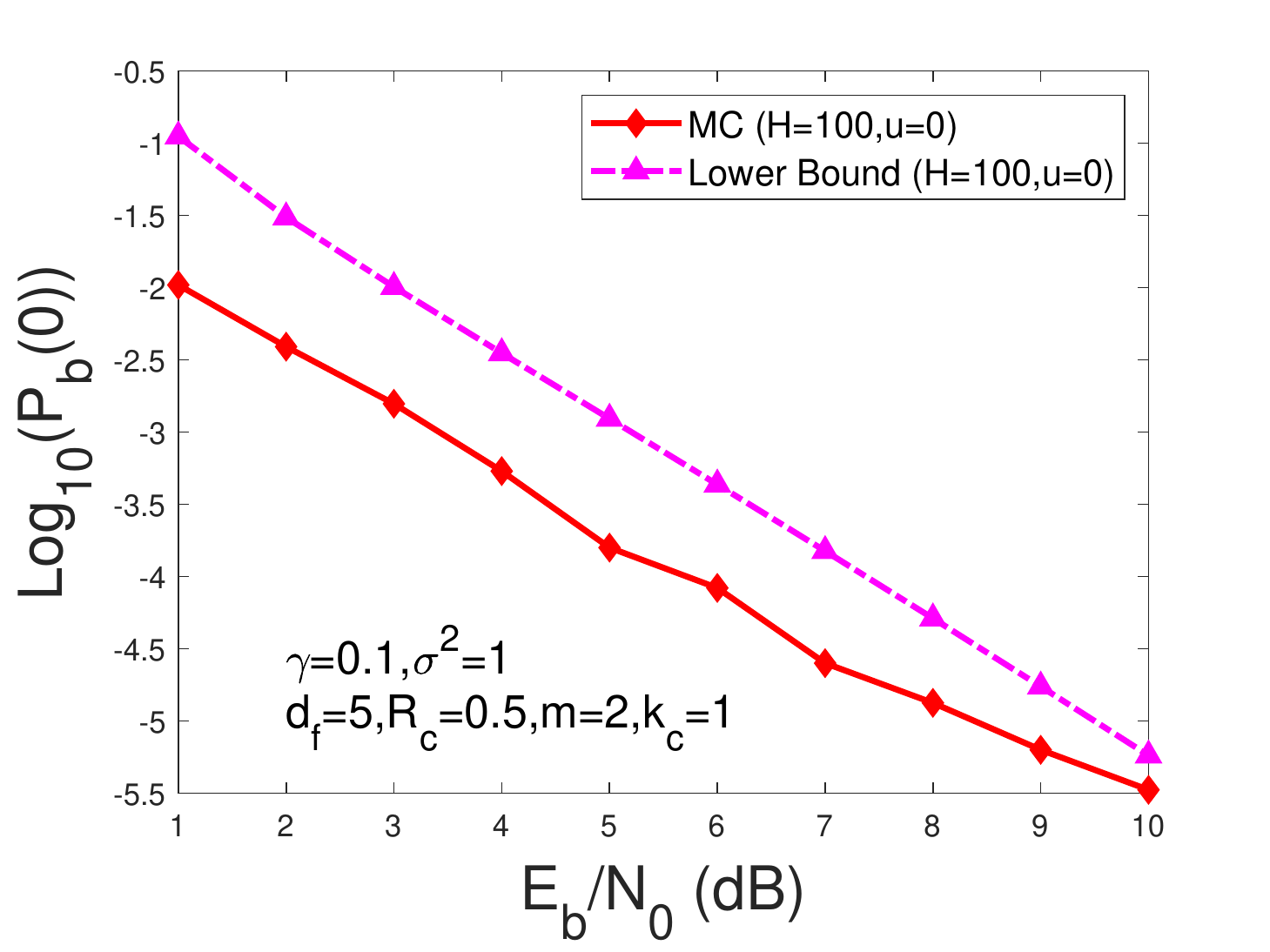} 
    \caption{Monte-Carlo vs. Bounds.}
		\label{fig2a}
  \end{minipage}
  \hfill
  \begin{minipage}[t]{0.48\textwidth}
    \includegraphics[width=\textwidth]{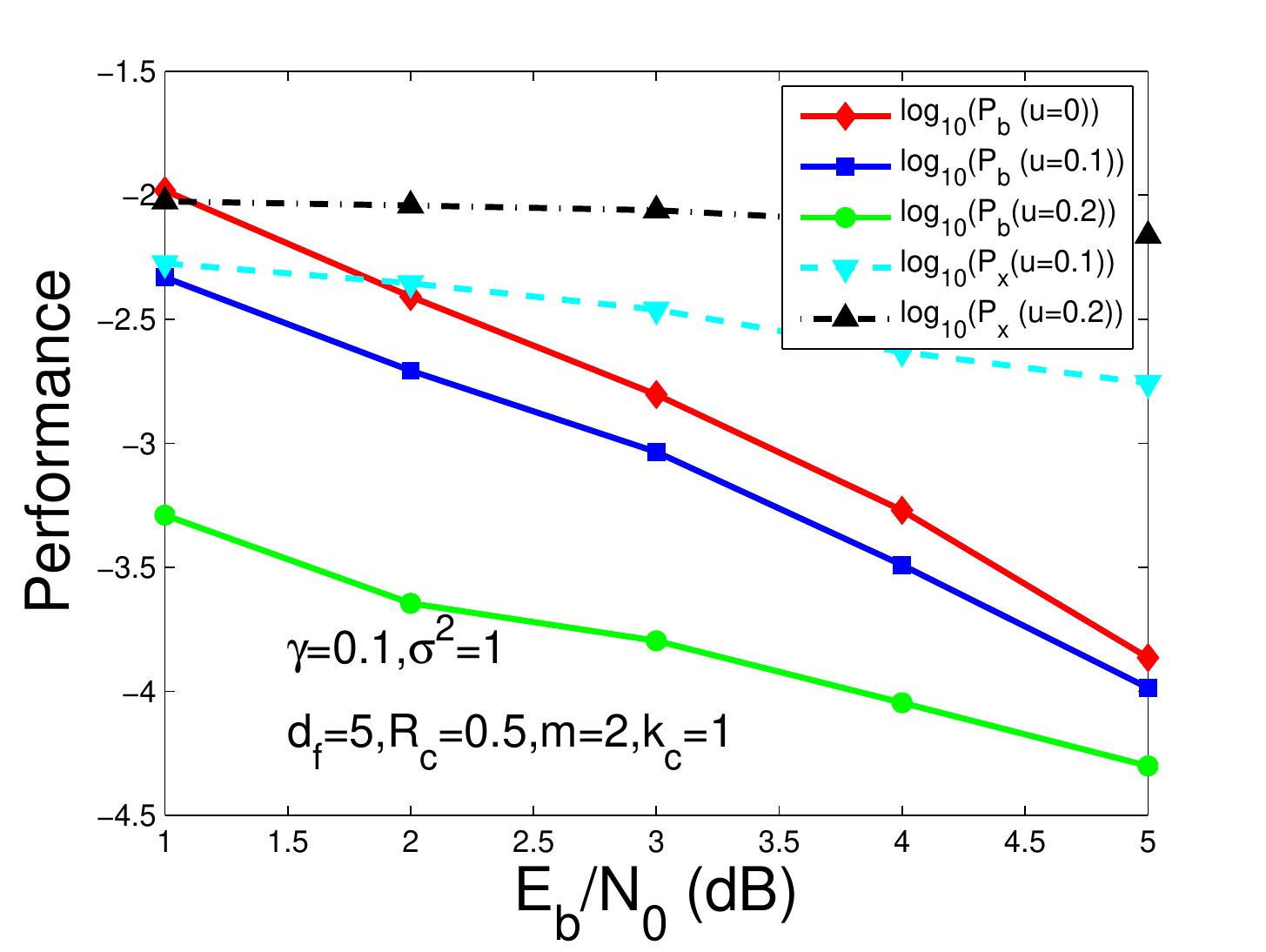} 
		\caption{Performance vs. $E_b/N_0$.}
    		\label{fig2b}
  \end{minipage}
\end{figure}
\begin{figure}[h] 
  \centering
  \begin{minipage}[t]{0.48\textwidth}
    \includegraphics[width=\textwidth]{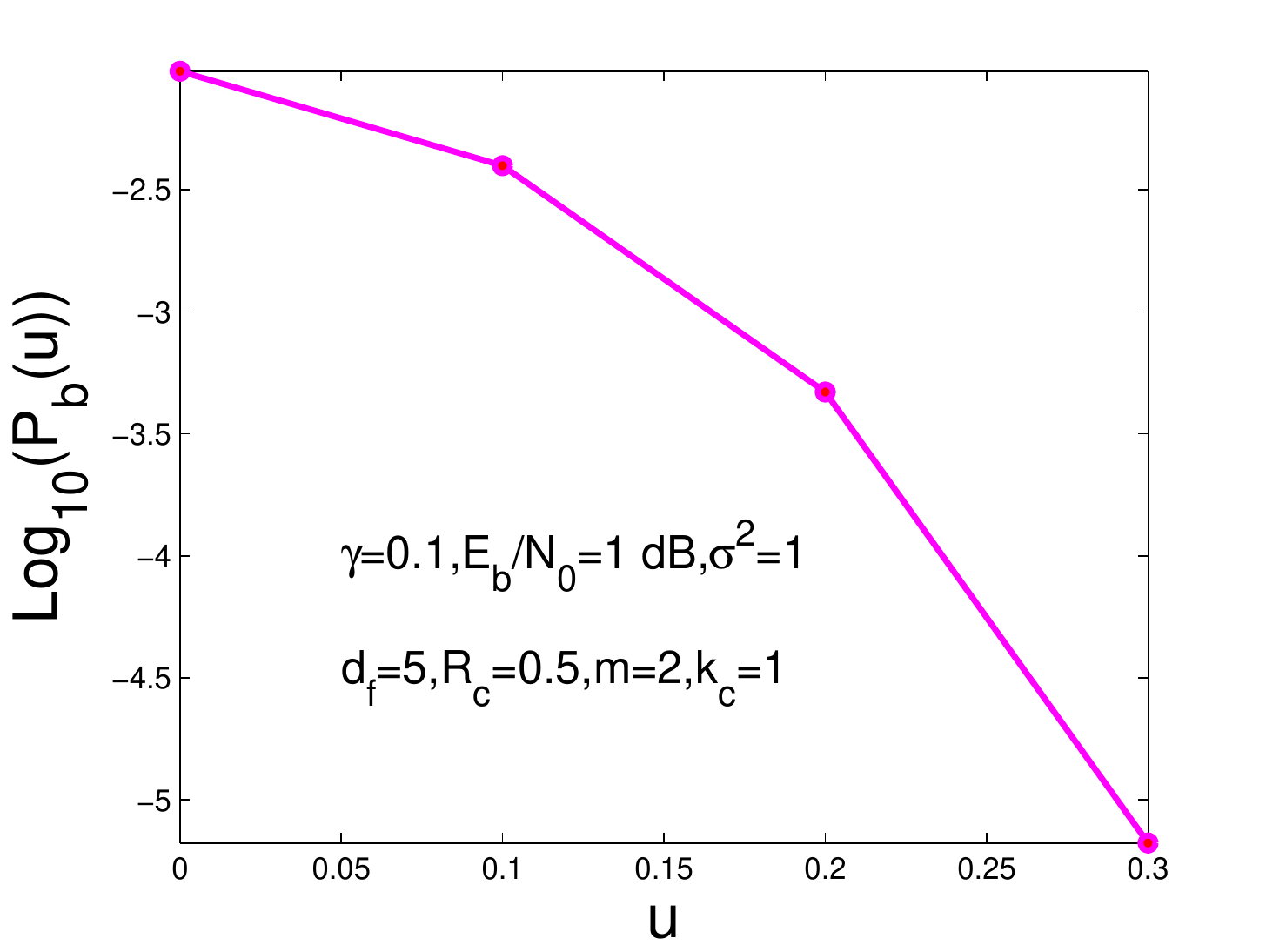} 
	  		\caption{BER vs. Yamamoto-Itoh Flag $(u)$.}
		\label{fig3a}
  \end{minipage}
	\hfill
   \begin{minipage}[t]{0.48\textwidth}
    \includegraphics[width=\textwidth]{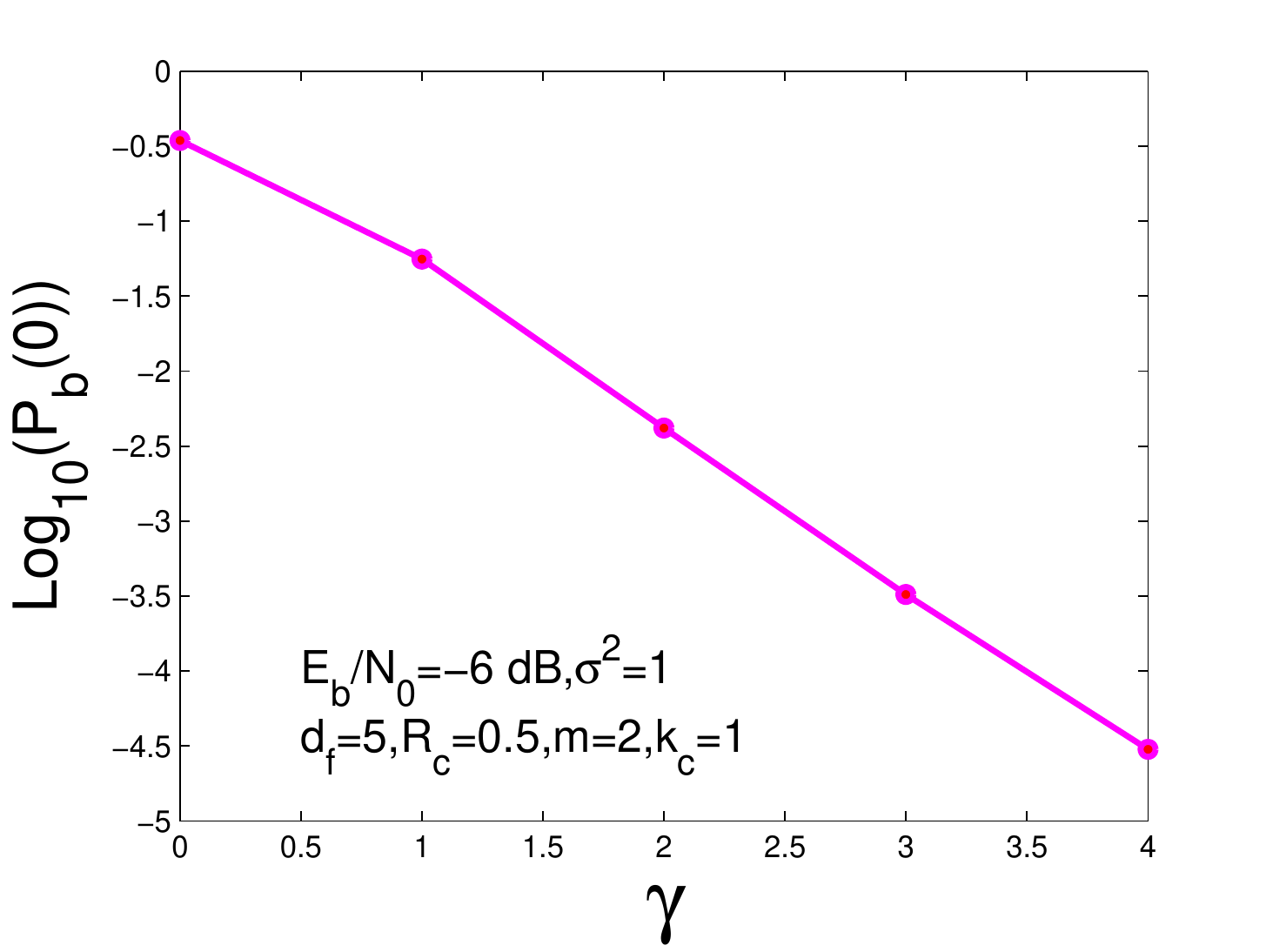}
    \caption{BER vs. Rician Factor $(\gamma)$.}
		\label{fig3b}
  \end{minipage}
\end{figure}
Let us consider a Monte Carlo simulation for the Viterbi coding scheme with $m=2\enspace (K=3), n_c=2, k_c=1, d_f=5$ in~\cite[Section II]{Viterbi1971}. This convolution coding scheme has the transfer function
\begin{align}
T(D,N)=\frac{D^5 N}{1-2DN}=\sum_{k=5}^{\infty} 2^{k-5} D^k N^{k-4},
\end{align} 
and
\begin{align}
\frac{d T(D,N)}{dN}=\sum_{k=5}^{\infty} 2^{k-5}(k-4) D^k N^{k-5}.
\end{align}
It follows from~\eqref{result1}--\eqref{result3} that
\begin{align}
\label{keyresult1}
\rvP_{\rme}(u)&\leq  \sum_{k=5}^{2(H+2)} 2^{k-4}\left[ \tilD(E_c/N_0, u/d_f,\sigma^2, s)\right]^k.
\end{align}
Similarly, for $u < 5 \sqrt{\frac{2E_c}{N_0}}$ we also obtain
\begin{align}
\label{keyresult2}
\rvP_{\rmx}(u) &\leq \sum_{k=5}^{2(H+2)} 2^{k-5} \left[ \tilD(E_c/N_0, -u/d_f,\sigma^2, s)\right]^k.
\end{align}
In addition, we also have
\begin{align}
\label{keyresult3}
\rvP_{\rmb}(u)& \leq \sum_{k=5}^{2(H+2)}  (k-4) 2^{k-5}\left[ \tilD(E_c/N_0, u/d_f,\sigma^2, s)\right]^k.
\end{align}
Here, $\tilD(E_c/N_0, u/d_f,\sigma^2, s)$ is defined as~\eqref{eq27key} of Proposition~\ref{prop1}. 

Figure~\ref{fig2a} compares Monte-Carlo simulation result of $\rvP_{\rmb}(0)$ with numerical evaluations of performance bounds in~\eqref{keyresult1} at $u=0$. The existing gap between associated curves can be explained as follows. The performance bounds in~\eqref{keyresult1} is based on an assumption that Hamming distance between the correct path and incorrect one at any merging node on the trellis diagram of the convolutional code can be in the range $[1, 2(H+2)]$, but this assumption looks not good enough. The Hamming distance range is actually dependent on the position of each merging node. Since positions of merging nodes are not easy to determine on trellis diagrams of convolutional codes, the method to evaluate convolutional code performance by using transfer function~\cite{Viterbi1971},~\cite{YamamotoItoh1980} usually creates loose bounds. However, as $E_b/N_0$ increases, the performance gap on this figure is narrowed down. Since Theorem~\ref{thm} considers asymptotic results, tight bounds as $E_b/N_0$ and/or $\gamma$ sufficiently large are good enough for results in this theorem to hold.

Figure~\ref{fig2b} shows $\rvP_{\rmb}(u)$ and~$\rvP_{\rmx}(u)$ as functions of bit energy to noise ratio $E_b/N_0$ for a fixed Rician factor $\gamma$ and for different values of Yamamoto-Itoh flag $u$. There is a tradeoff between the bit error probability $\rvP_{\rmb}(u)$ and the retransmission probability $\rvP_{\rmx}(u)$. As we increase Yamamoto-Itoh flag $u$,~$\rvP_{\rme}(u)$ decreases, but~$\rvP_{\rmx}(u)$ increases. 

Figure~\ref{fig3a} draws the bit error probability $\rvP_{\rmb}(u)$ as a function of $u$ for a fixed value of bit energy to noise ratio $E_b/N_0$ and a fixed value of Rician factor $\gamma$. The Viterbi decoding scheme using Yamamoto-Itoh flag helps to reduce the bit error probability $\rvP_{\rme}(u)$ compared with the Viterbi's decoding scheme without using this flag (i.e., $u=0$). 

Figure~\ref{fig3b} draws the the bit error probability of the original Viterbi decoding,~\cite{Viterbi1967}, $\rvP_{\rmb}(0)$ as a function of Rician factor $\gamma$ for a fixed value of $E_b/N_0$.
\section{Conclusion}
The performance of the Viterbi decoding algorithm with/without Automatic Repeat reQuest (ARQ) over a Rician flat fading channel with unlimited interleaving was evaluated. Our obtained results prove that the Rician factor causes the average bit error probability to have an exponential decay curve in simulations of Viterbi decoding performance in fading environments. In addition, the bit energy to noise ratio is shown to have an effect on the average bit error probability in a different way. More specifically, the average bit error exponent with respect to Rician factor is $d_f$ in unlimited interleaving fading environment, and the decay rate of the average bit error probability with respect to the bit energy to noise ratio is at least equal to $d_f$ in that environment. The Yamamoto-Itoh flag has been known to at least double the bit error exponent in DMCs, but the affects of Yamamoto-Itoh flags on the convolutional code performance in other channel models have been still open.  This paper shows an interesting fact that the Yamamoto-Itoh flag helps to reduce the average bit error probability by a factor of $4^{d_f}$. A sketch evaluation of lower bounds on performance of Viterbi decoding on interleaved Rician fading channels is given in~\cite{Truong18e}. 
\appendices
\section{Proof of Lemma \ref{lemeasy} }\label{app:prf_moments}
Observe that
\begin{align}
&\int_0^{z}   \exp\left(-\Phi_1 \alpha^2 -\Phi_2 \alpha\right) d\alpha\nonumber\\
&=\exp\left(\frac{\Phi_2^2}{4\Phi_1}\right) \int_0^{z} \exp\left[-\Phi_1\left(\alpha+\frac{\Phi_2}{2\Phi_1}\right)^2\right] d\alpha\\
&=\exp\left(\frac{\Phi_2^2}{4\Phi_1}\right) \int_{\Phi_2/(2\Phi_1)}^{z+\Phi_2/(2\Phi_1)} \exp(-\Phi_1\alpha^2) d\alpha\\
\label{smallkey}
&=\frac{\sqrt{2\pi}}{\sqrt{2\Phi_1}} \exp\left(\frac{\Phi_2^2}{4\Phi_1}\right) \nonumber\\
&\qquad \times \left[Q\left(\frac{\Phi_2}{\sqrt{2\Phi_1}}\right)-Q\left(\sqrt{2\Phi_1}\left[z+\frac{\Phi_2}{2\Phi_1}\right]\right)\right].
\end{align}
It follows from~\eqref{smallkey} that
\begin{align}
\label{newkey2}
&\varphi(\Phi_1,\Phi_2,z)=\int_0^{z} \alpha \exp\left(-\Phi_1 \alpha^2 -\Phi_2 \alpha\right) d\alpha\\
&=-\frac{1}{2\Phi_1} \exp\left(-\Phi_1 \alpha^2 -\Phi_2 \alpha\right)\Big|_0^{z} \nonumber\\
&\quad  -\frac{\Phi_2}{2\Phi_1}\int_0^{z} \exp\left(-\Phi_1 \alpha^2 -\Phi_2 \alpha\right) d\alpha \\
\label{newkey3}
&=\frac{1}{2\Phi_1}\bigg[1-\exp(-\Phi_1 z^2-\Phi_2 z)-\frac{\Phi_2\sqrt{2\pi}}{\sqrt{2\Phi_1}} \exp\left(\frac{\Phi_2^2}{4\Phi_1}\right) \nonumber\\
&\qquad \times \left[Q\left(\frac{\Phi_2}{\sqrt{2\Phi_1}}\right)-Q\left(\sqrt{2\Phi_1}\left[z+\frac{\Phi_2}{2\Phi_1}\right]\right)\right]\bigg].
\end{align}
Finally, we draw~\eqref{easyresult2} from~\eqref{newkey3} by taking $z\to \infty$.
\section{Proof of Lemma \ref{lem1} }\label{app:prf_lem1}
Observe that for each $k\in \{1,2,\ldots\}$, $a_k$ is equal to the number of paths of the Hamming distance $k$ from the all zeros, and $c_k$ is equal to the number of bits in error caused by an incorrect choice of the surviving path of free distance $k$ from the correct one~\cite{Viterbi1971}. Since convolutional codes are linear, each path of the Hamming distance $k$ from the correct one can be bijectively mapped to a path of the Hamming distance $k$ from all zeros. Since $k$ coded symbols in error can be at most in $k$ branches of the path, hence they can cause at most $k k_c$ bits to be in error. It follows that
\begin{align}
c_k \leq k k_c a_k. 
\end{align}

Now, we find the number of paths $a_k(L)$ of the Hamming weight $k$ and length $L$, which merge with the all zeros at a given node in a Viterbi trellis diagram. Observe that for each branch the Hamming distance between two paths is at most $n_c$, hence it is easy to see that 
\begin{align}
a_k(L)=0,\quad \forall k> Ln_c.
\end{align}
Now, for each $k\leq Ln_c$, we will show that
\begin{align}
\label{eq129:mod}
a_k(L)\leq \sum_{l=0}^{L-1}  {L \choose l} {k-1 \choose L-l-1}\bone\{L-l-1\leq k-1\}.
\end{align}
Indeed,~\eqref{eq129:mod} can be proved as follows. For a given path of length $L$, denote by $b_j$ the Hamming weight of branch $j$ for all $j=1,2,\cdots,L$. Then, each path of length $L$ with the Hamming distance $k$ from the all zeros corresponds to a tuple $(b_1,b_2,\cdots,b_L)$ which satisfies $\sum_{j=1}^L b_j=k$ where $0\leq b_j \leq n_c$. Note that the number of tuples $(b_1,b_2,\cdots,b_L)$ such that $\sum_{j=1}^L b_j=k$ is at most $\sum_{l=0}^{L-1}{L \choose l} {k-1 \choose L-l-1}$, which can be easily shown by using the following combinatorial choices.
\begin{itemize}
\item Choose $l$ out of $L$ numbers $(b_1,b_2,\cdots,b_L)$ and put them be equal to zero. There are ${L \choose l}$ ways of choices.
\item Find $L-l$ positive integers which sum up to $k$. 
\end{itemize}
To find $L-l$ positive integers which sum up to $k$, we put $k$ ones in a row, and then find $L-l-1$ positions to divide this row of $1$'s into $L-l$ parts where $b_j$ is equal to the number of $1$ in the part $j$ for each $j=1,2,\cdots,L$. There are ${k-1 \choose L-l-1} \bone\{L-l-1\leq k-1\}$ ways of dividing this row of ones into $L$ parts where each part is non-empty set. Hence, without constraining all the numbers less than or equal to $n_c$, the number of tuples of positive integers which sum up to $k$ is ${k-1 \choose L-l-1}\bone\{L-l-1\leq k-1\}$. Note that since we have a constraint $0\leq b_j \leq n_c$, hence the number of positive integer tuples which sum up to $k$ is at most ${k-1 \choose L-l-1}\bone\{L-l-1\leq k-1\}$. 

It follows that
\begin{align}
\label{eqak}
a_k &\leq \sum_{L=1}^{\lfloor \frac{k}{n_c} \rfloor} a_k(L) \\
\label{stack2}
&\leq \sum_{L=1}^{k} \sum_{l=0}^{L-1}{L \choose l} {k-1 \choose L-l-1}\bone\{L-l-1\leq k-1\}\\
\label{stack3}
&< \sum_{L=1}^{k} \sum_{l=0}^{L-1}{L \choose l} 2^{k-1} \\
\label{stack4}
&<\sum_{L=1}^{k} 2^L 2^{k-1}\\
&< 4^k.
\end{align}
Here,~\eqref{stack2} follows from the fact that $n_c\geq 1$ and~\eqref{eq129:mod},~\eqref{stack3} follows from the fact that 
\begin{align}
{k-1 \choose L-l-1}\bone\{L-l-1\leq k-1\}&<\sum_{t=0}^{k-1} {k-1\choose t}\\
&= 2^{k-1},
\end{align} and~\eqref{stack4} follows from the fact that $\sum_{l=0}^{L-1} {L \choose l} <\sum_{l=0}^L {L \choose l} =2^L$.
That concludes our proof of Lemma \ref{lemeasy}. 
\subsection*{Acknowledgements}
The authors are extremely grateful to the associate editor Prof.\ Alexandre Graell i Amat and the anonymous reviewers for their excellent and detailed comments that helped to correct typos, to remove imprecise, and to improve the readability of the paper. The author would also like to thank Prof.\ Vincent Y. F. Tan, Prof.\ Teng J. Lim (National University of Singapore), and Prof.\ Hirosuke Yamamoto (The University of Tokyo) for suggestions to improve the paper.  
\vspace{5mm}
\bibliographystyle{IEEEtran}
\bibliography{IEEEabrv,isitbib}
\begin{IEEEbiographynophoto}{Lan V. Truong} (S'12-M'15) received the B.S.E.\ degree in Electronics
and Telecommunications from Posts and Telecommunications Institute of
Technology (PTIT), Hanoi, Vietnam in 2003. After several years of working
as an operation and maintenance engineer (O\&M) at MobiFone Telecommunications
Corporation, Hanoi, Vietnam, he resumed his graduate studies at
School of Electrical \& Computer Engineering (ECE), Purdue University,
West Lafayette, IN, United States and got the M.S.E.\ degree in 2011. Then, he spent one year as a research assistant at NSF Center for Science of Information and Department of Computer Science, Purdue University in 2012. From 2013 to June 2015, he was an academic lecturer at Department of Information Technology Specialization (ITS), FPT University, Hanoi, Vietnam. From August 2015 to September 2018, he was a Ph.D. student at Department of Electrical \& Computer Engineering (ECE), National University of Singapore (NUS), Singapore. Since August 2018, he has been working as a Research Assistant/Postdoctoral Research Fellow at the Department of Computer Science, School of Computing, National University of Singapore. His research interests include information theory, communications, and machine learning.
\end{IEEEbiographynophoto}


\end{document}